\documentclass[%
 preprint,
 superscriptaddress,
 amsmath,amssymb,
 aps,
 showkeys,
 titlepage,
 endfloats*.   
]{revtex4-2}

\usepackage[utf8]{inputenc}
\usepackage[english]{babel}
\usepackage{amsmath}
\usepackage{tabularx,booktabs}

\usepackage{graphicx}
\usepackage{dcolumn}
\usepackage{bm}

\usepackage[colorlinks = true,
            linkcolor = blue,
            urlcolor  = blue,
            citecolor = red,
            anchorcolor = blue]{hyperref}

\begin{document}


\title{Impact of Dimensionality on the Magnetocaloric Effect in Two-dimensional Magnets} 

\author{Lokanath Patra}
\affiliation{Department of Mechanical Engineering, University of California, Santa Barbara, CA 93106, USA}

\author{Yujie Quan}
\affiliation{Department of Mechanical Engineering, University of California, Santa Barbara, CA 93106, USA}

\author{Bolin Liao}
\email{bliao@ucsb.edu} \affiliation{Department of Mechanical Engineering, University of California, Santa Barbara, CA 93106, USA}

\date{\today}

\begin{abstract}
Magnetocaloric materials, which exploit reversible temperature changes induced by magnetic field variations, are promising for advancing energy-efficient cooling technologies. The potential integration of two-dimensional materials into magnetocaloric systems represents an emerging opportunity to enhance the magnetocaloric cooling efficiency. In this study, we use atomistic spin dynamics simulations based on first-principles parameters to systematically evaluate how magnetocaloric properties transition from three-dimensional (3D) to two-dimensional (2D) ferromagnetic materials. We find that 2D features such as reduced Curie temperature, sharper magnetic transition, and higher magnetic susceptibility are beneficial for magnetocaloric applications, while the relatively higher lattice heat capacity in 2D can compromise achievable adiabatic temperature changes. We further propose GdSi$_2$ as a promising 2D magnetocaloric material near hydrogen liquefaction temperature. Our analysis offers valuable theoretical insights into the magnetocaloric effect in 2D ferromagnets and demonstrates that 2D ferromagnets hold promise for cooling and thermal management applications in compact and miniaturized nanodevices.
\end{abstract}

                            
\maketitle


\section{Introduction}
Over the past few years, there has been significant interest in the exploration of novel magnetic properties and their applications in two-dimensional (2D) magnets.~\cite{gong2017discovery, huang2017layer, bonilla2018strong, deng2018gate, tan2018hard, kong2019vi3} These magnets have sparked numerous investigations, particularly in the context of their monolayer form.~\cite{lado2017origin, chen2019boosting, pizzochero2020magnetic, lin2018critical, huang2018electrical, niu2019coexistence} The emergence of additional spin degrees of freedom compared to nonmagnetic 2D materials has presented fascinating possibilities in fields such as spintronics~\cite{wang2020electrically} and magnetic memory devices.~\cite{watanabe2018shape,hou20192d, zhang2020memory}. Fundamentally, these applications are enabled by several unique physical characteristics of 2D magnets compared to their 3D counterparts. For example, in van der Waals (vdW) layered lattices, a notable feature is the existence of strong anisotropy in exchange interactions, with varying strengths along distinct spatial directions.~\cite{jiang2021recent} In addition, 2D magnets exhibit conventional exchange interactions along with novel and complex exchange interaction mechanisms such as super-superexchange, extended superexchange, and multi-intermediate double exchange interactions.~\cite{yu2022recent} The existence and interplay of these various exchange mechanisms contribute to the rich and diverse landscape of magnetic interactions in 2D magnets. Furthermore, in contrast to their bulk counterparts, 2D magnetism exhibits strong critical fluctuations that are sensitive to external perturbations.~\cite{jin2020imaging} Practically, 2D magnets can be flexibly integrated into heterostructures, show pronounced quantum confinement, and provide full tunability through an external electric field or strain, showcasing the significant potential for voltage- or strain-controlled magnetic applications. 

Compared to the extensive efforts focused on 2D magnets for information technology~\cite{song2018giant, frisenda2018recent, gibertini2019magnetic, burch2018magnetism} and spintronics,~\cite{chumak2015magnon, macneill2017control, sadovnikov2017spin, zhang2019van} the study of other characteristic properties of 2D magnets and their applications remains relatively underdeveloped. One example is the magnetocaloric effect in 2D magnets and its potential application in magnetic cooling technologies. The technology of magnetocaloric cooling is based on the magnetocaloric effect (MCE).~\cite{sokolovskiy2022review, gomez2013magnetocaloric, franco2012magnetocaloric} By subjecting magnets to an external magnetic field, it is possible to achieve a reversible change in temperature and entropy due to the alignment of the magnetic moments with the external magnetic field. A cooling cycle can be constructed based on these processes.~\cite{lyubina2017magnetocaloric} Bulk magnetocaloric materials have been widely used for cooling applications in space missions, astrophysics, low-temperature scientific experimentation, and as an environmentally friendly alternative to vapor compression cycles.~\cite{omer2008energy, shen2009recent, gutfleisch2011magnetic, gutfleisch2016mastering, li2020understanding} 

In contrast, research on the MCE in layered van der Waals (vdW) materials and their 2D monolayer limit remains scarce. Recent studies have focused on investigating the MCE in bulk-layered vdW magnets. These studies have reported figures of merit for MCE evaluation, such as the isothermal magnetic entropy change ($\Delta S_{M}$) and the adiabatic temperature change ($\Delta T_{ad}$).~\cite{liu2018anisotropic, yu2019large, liu2020anisotropic, mondal2020magnetic, liu2020critical, tran2022insight} $\Delta S_{M}$ and $\Delta T_{ad}$ can be determined by analyzing the magnetization in relation to temperature and applied magnetic field, using thermodynamic Maxwell relations: ~\cite{pecharsky1999magnetocaloric,patra2023indirect}
\begin{equation}
\Delta S_{M} (T, \Delta H) = \int_{H_i}^{H_f} \left( \frac{\partial M(T,H)}{\partial T} \right)_H\ dH,
\label{eqn:entropy_change}
\end{equation} and
\begin{equation}
\Delta T_{ad} (T, \Delta H) = \int_{H_i}^{H_f} \frac{T}{C_p} \left( \frac{\partial M(T,H)}{\partial T} \right)_H\ dH,
\label{eqn:T_change}
\end{equation}
respectively, where $\Delta H = H_f - H_i$ is the change of the applied external field ($H_f$ and $H_i$ are final and initial fields, respectively), $C_p$ is the specific heat capacity under a constant pressure, $M$ is the magnetization, and $T$ is the temperature. For example, MCE properties of the bulk crystals of the layered chromium halide family (CrI$_3$, CrBr$_3$, and CrCl$_3$) have recently been studied experimentally, showing great promise for practical applications particularly at cryogenic temperatures.~\cite{liu2018anisotropic, tran2022insight,yu2019large,liu2020anisotropic,mondal2020magnetic} However, it remains unclear whether the weak interlayer magnetic coupling contributes to their high MCE performance. 

Furthermore, only a very limited number of studies have examined how the MCE properties differ in the bulk and the monolayer limit. Recently, He et al.~\cite{he2023giant} reported a giant MCE and its strain tunability in monolayer magnets, including CrX$_3$ (X = F, Cl, Br, and I) and CrAX (A = O, S, and Se) compounds. Notably, CrF$_3$ exhibits excellent MCE (maximum $-\Delta S_{M} \sim 35\,\mu J\,m^{-2} K^{-1}$ and $\Delta T_{ad} \sim~11$\, K) at low temperatures, while CrOF and CrOCl excel at medium temperatures. They further showed that MCE in these 2D magnets can be enhanced by a compressive strain. Despite these promising results, a fundamental understanding of how the reduced dimensionality impacts the magnetocaloric properties of 2D magnets is still lacking. For example, from Eqns.~\ref{eqn:entropy_change} and \ref{eqn:T_change}, $\Delta S_{M}$ and $\Delta T_{ad}$ sensitively depend on the saturation magnetization, the Curie temperature, how the magnetization changes with temperature near the Curie temperature, and the specific heat capacity, all of which are qualitatively impacted by the reduced dimensionality in 2D magnets.

 In this work, we investigate the MCE response in 2D magnets by employing atomistic spin dynamics (ASD) simulation with exchange interaction parameters evaluated from first-principles density functional theory (DFT) calculations. In particular, we systematically examine the influence of critical factors such as the weakening of interlayer exchange interactions, the Curie temperature (T$_{C}$), field-dependent saturation magnetization, and specific heat capacity, revealing the mechanisms driving the difference between 2D and 3D bulk materials in terms of the MCE. Through detailed studies of the MCE in model systems such as CrX$_3$ (X = F, Cl, Br, I) and GdSi$_2$ monolayers, we show that 2D magnets exhibit a lower T$_c$ and sharper transition of magnetization with temperature near T$_c$ due to weakened interlayer vdW interactions and stronger critical fluctuations. These are beneficial for obtaining a higher $\Delta S_{M}$ and indicate that 2D magnetocalorics can potentially be driven by a smaller external magnetic field. We also show that, although the magnetic specific heat is suppressed in 2D magnets compared to 3D bulk materials, the phononic specific heat is generally enhanced due to the flexural modes, limiting the enhancement in $\Delta T_{ad}$. Our work provides a systematic physical framework to understand the characteristic behaviors of the magnetocaloric effect in 2D magnets. Our work also suggests GdSi$_2$ as a novel 2D magnet with promising magnetocaloric properties for applications near 20\,K, such as hydrogen liquefaction.~\cite{numazawa2014magnetic}       

\section{Computational Details}

Density functional theory (DFT) calculations were performed using the Vienna ab initio simulation package (VASP), employing projected augmented wave pseudopotentials.~\cite{kresse1996efficient, blochl1994projector} Structural optimization was carried out using the Perdew-Burke-Ernzerhof form of the generalized gradient approximation (PBE-GGA).~\cite{perdew1996generalized} A plane-wave cut-off energy of 600 eV was utilized for all calculations. We set the energy convergence criterion to 1 $\times$ 10$^{-5}$ eV and the force convergence criterion to 0.01 eV/\AA. For the structural optimization of CrX$_3$ (X = F, Cl, Br, I) and GdSi$_2$, we employed Monkhorst-Pack \textbf{k}-point grids~\cite{pack1977special} of 8 $\times$ 8 $\times$ 1 to sample the Brillouin zone. For accurately determining the energy differences between different magnetic configurations, a denser \textbf{k}-grid of 12 $\times$ 12 $\times$ 1 was used. 

ASD simulations were performed with UppASD code,~\cite{skubic2008method} employing magnetic exchange energies obtained from DFT calculations. The Monte Carlo method was utilized to estimate the $T_C$. This involves a system of $30 \times 30 \times 1$ unit cells running for 10,000 steps at each temperature, enabling the in-plane periodic boundary conditions to reach equilibrium. Next, the mean magnetization was extracted by a statistical average obtained over 10,000 steps. Temperature-dependent magnetization (M $\sim$ T) curves were generated using the spin dynamics approach, while systematically increasing the external magnetic field to 5\,T with a step size of 1\,T. Using these M $\sim$ T curves, $T_C$ values were estimated by fitting the curves with $M(T) = \left(1 - \frac{T}{T_C}\right)^\beta$, where $\beta$ denotes the critical exponent. The field-dependent M $\sim$ T curves are utilized to evaluate $\Delta S_M$  using Equation~\ref{eqn:entropy_change}.

The total specific heat ($C_T$) is another essential parameter for estimating the MCE performance of materials and is typically determined as a combination of phononic ($C_p$), magnonic ($C_m$), and electronic ($C_e$) contributions, expressed as:

\begin{equation}
    C_T = C_m + C_p + C_e.
\end{equation}

However, since the CrX3 monolayers are semiconductors, we can neglect the electronic contribution ($C_e$), simplifying the expression to:

\begin{equation}
    C_T = C_m + C_p. 
\end{equation}

 We start by determining the magnetic energy (U) of the magnetic compound to derive $C_m$ i.e.
 \begin{equation}
     U = -\sum_{ij}\lambda_{ij}\Vec{J_i} \cdot \Vec{J_j}-\sum_{i}g\mu_B\Vec{J_i}\cdot h^{ext}.
 \end{equation}
In this equation, the first term indicates the spin-spin exchange interaction, where $\lambda_{ij}$ is the exchange interaction parameter and $\Vec{J_i}$ denotes the total angular momentum of the magnetic ions. The second term represents the Zeeman interaction of the total angular momentum with the external applied field $h^{ext}$. To calculate the mean energy value $\langle U \rangle$ at a given temperature, we employ 
 \begin{equation}
     \langle U \rangle = \frac{1}{(N_C-N_0)}\sum_{i>N_0}^{N_C}U_i,
     \label{eqn-U-mean}
 \end{equation}
where $N_C$ and $N_0$ represent the total number of Monte Carlo cycles and the number of cycles used for thermalization, respectively. The mean square energy $\langle U^2 \rangle$ is obtained similarly to Equation~\ref{eqn-U-mean}. Now, the magnetic heat capacity is calculated as
\begin{equation}
    C_m(T,h^{ext})=\frac{\langle U^2 \rangle - \langle U \rangle^2}{k_BT^2},
\end{equation}
where $k_B$ is the Boltzmann constant. The simulations were performed using the UppASD software,~\cite{skubic2008method} which provides the magnetic heat capacity as one of the outputs. 

After determining the phonon frequencies via the frozen phonon method,  we calculate the constant volume phonon heat capacity ($C_V$) using the equation implemented in the Phonopy code~\cite{phonopy-phono3py-JPCM,phonopy-phono3py-JPSJ}: 

\begin{equation}
    C_V = \sum_{qj} C_{qj} = \sum_{qj} k_B \left(\frac{\hbar \omega_{qj}}{k_BT}\right)^2 \frac{\exp (\hbar \omega_{qj}/k_BT)}{[\exp (\hbar \omega_{qj}/k_BT)-1]^2},
\end{equation}
where $q$ represents the wave vector, $j$ is the band index, $\omega_{qj}$ denotes the phonon frequency, and $\hbar$ is the reduced Planck's constant. We approximate calculated $C_V$ as the constant-pressure heat capacity $C_p$, as in the low-temperature regime of solids, $C_V \approx C_p$.~\cite{wang2012magnetic} Next, we estimate $\Delta T_{ad}$ employing the calculated $C_T$ in Equation~\ref{eqn:T_change}:

\section{Results and Discussions}
\subsection{Curie Temperature and Temperature-dependent Magnetization}
\begin{figure}[!tb]
\includegraphics[width=\textwidth]{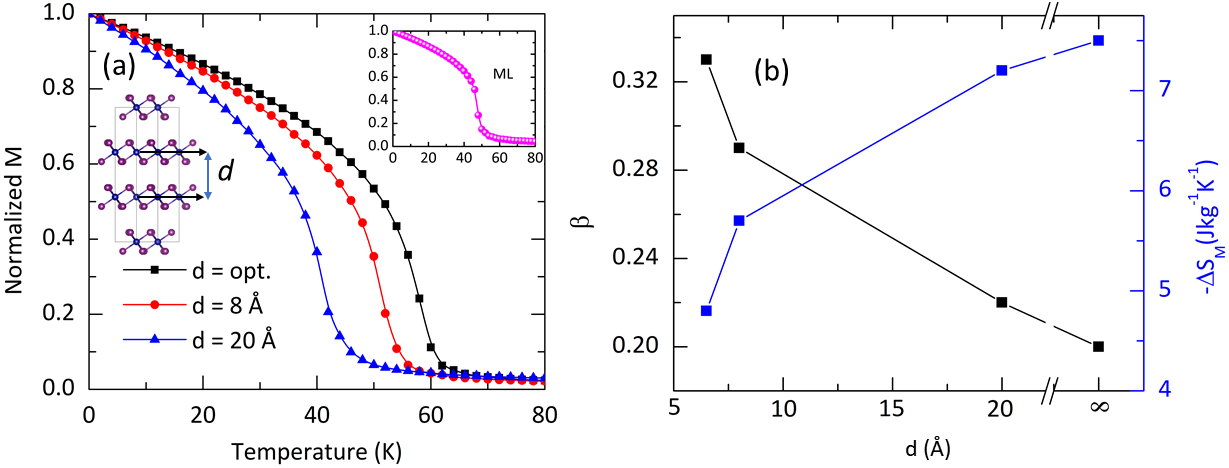}
\caption{\textbf{Transition of the magnetocaloric properties from 3D to 2D in CrI$_3$ by increasing the interlayer distance.} (a) Magnetization vs. temperature ($M \sim T$) curves for CrI$_3$ with different interlayer distance ($d$). The inset shows the atomic structure of CrI$_3$. The optimized $d$ value of 6.5 \AA corresponds to the bulk CrI$_3$. (b) The critical exponent ($\beta$) and the magnetic entropy change ($\Delta S_M$) values as a function of $d$ for CrI$_3$. The M $\sim$ T curve for free-standing CrI$_3$ monolayer (ML) is also given as an inset of (a).}
\label{fig:fig1}
\end{figure}

First, we analyze how the transition from 3D to 2D affects the Curie temperature and the temperature derivative of the magnetization at the Curie temperature. Both parameters have a strong impact on the magnetocaloric entropy change $\Delta S_M$. For this purpose, we choose the CrX$_3$ family and simulate how their properties depend on the distance between adjacent atomic layers. The general structural configuration of bulk CrX$_3$ systems is depicted in the inset of Fig.~\ref{fig:fig1} (a), wherein six halide atoms are chemically bonded to a central Cr atom, forming polyhedral units. The crystalline arrangement of CrX$_3$ adopts a layered structure within a rhombohedral lattice framework (space group R$\overline{3}$). Table~\ref{tab:crx3} provides the calculated magnetocrystalline anisotropy energy (MAE) values for CrX$_3$ monolayers, which are in qualitative agreement with previous calculations.~\cite{he2023giant} Notably, the CrI$_3$ monolayer exhibits a substantial MAE of 0.643 meV, attributable to the pronounced influence of spin-orbit coupling (SOC) within the iodine ions. In the case of CrX$_3$, the out-of-plane MAE energy is lower than the in-plane counterpart, thus establishing an easy axis orientation along the out-of-plane $c$ direction.  Table~\ref{tab:crx3} also presents the exchange interaction parameters ($J_i$, where $i$ designates the $i^{th}$ nearest neighbor). Here positive values of $J$ denote a ferromagnetic coupling within the monolayer structures. In the context of bulk CrX$_3$, the strength of the out-of-plane exchange coupling interaction, denoted as $J_z$, follows the hierarchy $J_1$ $>$ $J_z$ $>$ $J_2$ $>$ $J_3$. This result is in line with the experimental observation, where the out-of-plane interaction was found to be comparable to the in-plane interaction in bulk CrI$_3$.~\cite{mcguire2015coupling} The dominant source of the robust J$_z$ coupling stems from the out-of-plane superexchange interaction.~\cite{whangbo2003spin} J$_z$ gradually diminishes to a negligible value with an increase in the interlayer separation distance, denoted as $d$ in this work.

\begin{table}[!t]
    \centering
    \setlength{\extrarowheight}{5pt}
    \caption{The lattice constants, exchange energies ($J_1, J_2, J_3$), magnetic anisotropy energies (MAE), Curie temperatures ($T_C$), and magnetic entropy changes ($\Delta S_M$) for CrX$_3$ monolayers.}
    \label{tab:crx3}
    \begin{tabularx}{0.75\textwidth}{@{}lccccc@{}}
        \toprule
        System & Lattice constant  & $J_1, J_2, J_3$ & MAE (meV) & $T_C$ & $-\Delta S_M$  \\
               & (Å) & (meV) & (meV) & (K) & (J.kg$^{-1}$.K$^{-1}$) \\
        \midrule
        CrF$_3$ & 5.19 & 1.5, 0.06, -0.01 & 0.085 & 18 & 32.2 \\
        CrCl$_3$ & 6.06 & 1.7, 0.3, 0.01 & 0.035 & 22 & 21.9 \\
        CrBr$_3$ & 6.44 & 2.4, 0.5, 0.02 & 0.147 & 36 & 12.5 \\
        CrI$_3$ & 7.01 & 2.7, 0.7, 0.04 & 0.643 & 48 & 7.1 \\
        \bottomrule
    \end{tabularx}
\end{table}

Figure~\ref{fig:fig1}(a) presents the magnetization versus temperature ($M$ vs. $T$) curves for various $d$ values in CrI$_3$ [d = optimized (6.5), 8, and 20 \AA]. In our analysis, the system with an optimized $d$ value represents the bulk configuration, while the system with $d$ = 20 \AA~ has negligible interlayer interactions, thus approaching a 2D limit. The calculated $T_C$ values are as follows: 58 K for bulk CrI$_3$, 43 K for the 2D system, and an intermediate value of 51 K for $d$ = 7 \AA.  Notably, the simulated $T_C$ value decreases as the $d$ value increases. This behavior is consistent with the expected trend, as the increasing distance between adjacent layers weakens J$_z$. We have also performed similar ASD calculations for the freestanding monolayer CrI$_3$ using the optimized lattice constants. The T$_C$ value was estimated to be 48 K, in good agreement with the reported measured value of 45 K.~\cite{huang2017layer} In general, a lower T$_C$ gives rise to a higher $\Delta S_M$ due to a stronger dependence of magnetization on temperature.

Next, we analyze the temperature-dependent magnetization in CrI$_3$ as a function of $d$ and estimate the critical exponents ($\beta$) for each case. Determining critical exponents provides valuable insights into the interactions occurring during the magnetic transition. It also directly affects the magnetocaloric entropy change, as $\Delta S_M$ is determined by the ``sharpness'' of the magnetization transition at the Curie temperature as suggested by Eqn.~\ref{eqn:entropy_change}. Taroni et al.~\cite{taroni2008universal} conducted a thorough examination of critical exponents and established a typical range of approximately $0.1 \le \beta \le 0.25$ for critical exponent $\beta$ in the context of 2D magnets (2D Ising model gives a critical exponent of 0.125).~\cite{gibertini2019magnetic} In contrast, 3D models provide a critical exponent ranging from 0.3265 to 0.369.~\cite{gibertini2019magnetic} Since a smaller critical exponent suggests a sharper transition of the magnetization, the smaller critical exponent associated with 2D magnets should lead to a higher $\Delta S_M$. To verify this hypothesis, we apply the Curie-Weiss function, $M \sim (1-\frac{T}{T_C})^\beta$ near the Curie temperature, to the magnetization data obtained from ASD simulations conducted on CrI$_3$ with different $d$ values [Fig.~\ref{fig:fig1}(a)]. The estimated $\beta$ values are determined to be 0.34 for the bulk and 0.22 for the $d$ = 20 \AA ~configuration [Fig.~\ref{fig:fig1}(b)], in good agreement with the expectation from analytical models.~\cite{gibertini2019magnetic} The continuously weaker out-of-plane interaction as a function of $d$ makes the out-of-plane magnetic moment alignment unstable, even at temperatures below the Curie temperature, leading to a reduced value of $\beta$.~\cite{tiwari2021critical} The $\beta$ value for the monolayer system is estimated to be 0.20. These values imply that the magnetic interactions within CrI$_3$ exhibit a noticeable transition from 3D to 2D behavior as a function of $d$. This observation highlights the critical role played by interlayer coupling in influencing the magnetic properties of this class of materials.

In the subsequent analysis, we verify that $\Delta S_M$ is enhanced due to the decreased $\beta$ during the 3D to 2D transition. The calculated $\Delta S_M$ values (with an applied magnetic field of 5 T) as a function of $d$ are given in Fig.~\ref{fig:fig1}(b). For bulk CrI$_3$, the calculated $\Delta S_M$ value is $\sim$ -4.75 J kg$^{-1}$ K$^{-1}$. The notable overestimation of the computed $\Delta S_M$ value in comparison to experimental data ($\sim$ -3.35 J kg$^{-1}$ K$^{-1}$)~\cite{tran2022insight} could be ascribed to disparities in sample preparation techniques and sample purity. Remarkably, there is an approximately 58\% enhancement in the $\Delta S_M$ value when transitioning from bulk to monolayer CrI$_3$. The computed value of $\Delta S_M$ for monolayer CrI$_3$ is $\sim$ -22 $\mu$J m$^{-2}$ K$^{-1}$ (equivalent to $\sim$ -7.5 J kg$^{-1}$ K$^{-1}$) under an applied magnetic field of 5 T [See Fig.~\ref{fig:fig1}(b)]. This enhancement is in qualitative agreement with the previous report by He et al.~\cite{he2023giant} The enhancement in the $\Delta S_M$ values observed during the transition from bulk to the 2D limit can be attributed to two factors: (i) a lower $T_C$ value due to the weaker out-of-plane exchange interaction, and (ii) a lower $\beta$ value indicating a sharper transition in the $M \sim T$ curves, resulting in a higher $\frac{\partial M}{\partial T}$. 

\subsection{Sensitivity of the MCE to External Magnetic Field}
\begin{figure}[!tb]
\includegraphics[width=\textwidth]{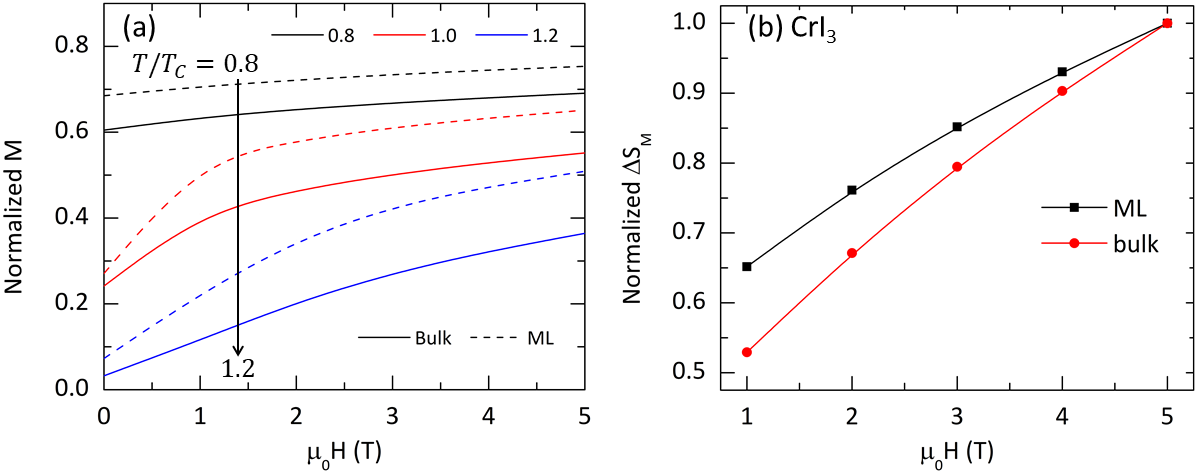}
\caption{\textbf{Field-dependent magnetic and magnetocaloric properties of bulk and monolayer CrI$_3$.} The normalized magnetization as a function of applied field for three temperatures around T$_C$ i.e. 0.8T$_C$, 1.0T$_C$, and 1.2T$_C$ for both bulk and monolayer CrI$_3$ systems. (b) The normalized magnetic entropy change ($\Delta S_M$) values as a function of the applied field for bulk and monolayer CrI$_3$; the monolayer shows higher values even at lower applied fields.}
\label{fig:fig2}
\end{figure}

Another important factor in MCE applications is the magnetic field required to drive the MCE cycle. The sensitivity of magnetization to external magnetic fields has been observed to be significantly higher in 2D ferromagnetic materials as compared to their bulk counterparts near the Curie temperature.~\cite{tokmachev2021two, bruno1991magnetization, bruno1991spin} In the former, even minute magnetic fields have demonstrated the capacity to induce a substantial increase in magnetization, consequently facilitating the attainment of saturation magnetization at considerably lower field strengths. This is due to stronger critical fluctuations near the magnetic transition in 2D magnets,~\cite{jin2020imaging} which leads to a larger critical exponent $\gamma$ for the divergence of the magnetic susceptibility ($\chi \sim |1-\frac{T}{T_C}|^{-\gamma}$) near the magnetic transition in 2D compared to 3D: 2D Ising model predicts $\gamma = 1.75$ while 3D models predict $\gamma$ in the range of 1.237 to 1.396.~\cite{gibertini2019magnetic} To evaluate this effect on the MCE in 2D magnets, Fig.~\ref{fig:fig2} (a) shows the simulated isothermal magnetization curves for both bulk and monolayer configurations of CrI$_3$, spanning various temperatures around the $T_C$, i.e. 0.8T$_C$, 1.0T$_C$, and 1.2T$_C$. A notable difference in saturation magnetization is observed when comparing the characteristics of bulk and monolayer CrI$_3$. Specifically, the magnetization was found to be higher in the monolayer configuration at a given field. Across all temperature ranges, the monolayer exhibited a faster approach to saturation magnetization at lower applied magnetic fields compared to the bulk configuration. In Fig.~\ref{fig:fig2}(b), we compare the $\Delta S_M$ values for bulk and monolayer CrI$_3$ at various applied fields at $T_C$. Notably, even at lower applied fields, the CrI$_3$ monolayer exhibits higher $\Delta S_M$ values compared to the bulk configuration. This suggests that higher $\Delta S_M$ values can be achieved with relatively lower external fields in a monolayer of vdW layered ferromagnets, another potential advantage of 2D magnets for MCE applications. 

\begin{figure}[!htb]
\includegraphics[width=\textwidth]{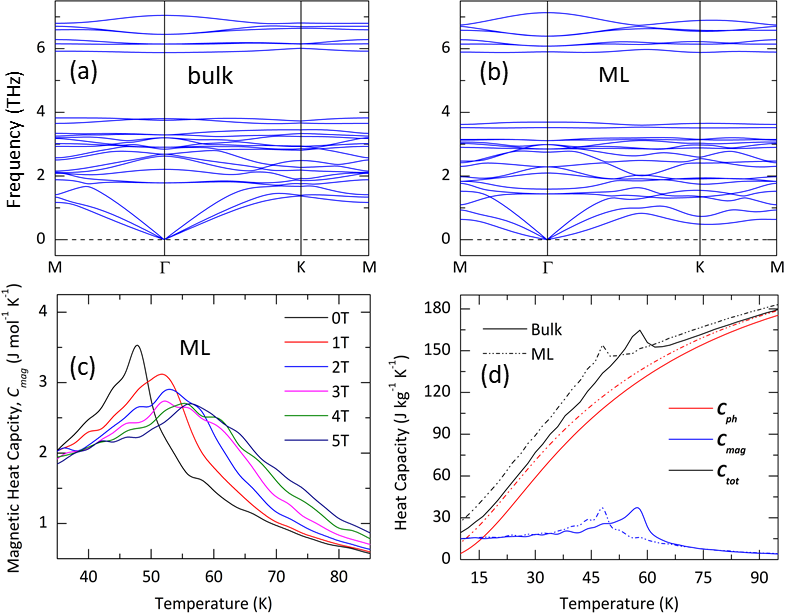}
\caption{\textbf{Heat capacities of bulk vs. monolayer configurations of CrI$_3$:} Phonon dispersions for CrI$_3$ are shown in (a) for the bulk and (b) for the monolayer configurations. The field-dependent magnetic heat capacity ($C_{mag}$) for the CrI$_3$ monolayer is shown in  (c). In (d), the total heat capacities ($C_{tot}$) are presented, with contributions from phonons ($C_{ph}$) and magnetism ($C_{mag}$), expressed as $C_{tot} = C_{ph} + C_{mag}$.}
\label{fig:fig3}
\end{figure} 

\subsection{Heat Capacity and Adiabatic Temperature Change}

In addition to $\Delta S_M$, we also evaluate the impact of dimensionality on the adiabatic temperature change $\Delta T_{ad}$. From Eqn.~\ref{eqn:T_change}, a larger $\Delta T_{ad}$ requires a lower total heat capacity. To understand how $\Delta T_{ad}$ evolves from bulk to the 2D limit, we simulate the magnetic-field-dependent specific heat for both bulk and monolayer CrI$_3$ to investigate the dimensionality effect. The total specific heat capacity of a magnetic material typically comprises lattice, electronic, and magnetic contributions. In semiconducting materials, the electronic contribution is negligible due to the absence of free electrons, thus omitted in our specific heat estimation. Figures~\ref{fig:fig3}(a) and (b) display phonon dispersions for bulk and monolayer CrI$_3$, affirming their dynamic stability (no imaginary frequencies are observed). Slightly softened phonons are observed in 2D due to the reduced interlayer interaction. Additionally, the out-of-plane flexural phonon mode with a quadratic dispersion emerges in the 2D limit. Overall lower phonon frequencies (and thus, a lower Debye temperature $T_D$) in 2D lead to a higher lattice heat capacity compared to that in 3D. Total heat capacities including lattice and magnetic contributions for both bulk and monolayer forms are presented in Fig.~\ref{fig:fig3}(c). Notably, heat capacity exhibits sharp discontinuities at 58 K and 48 K for bulk and monolayer CrI$_3$, respectively, due to magnetic transitions. Our analysis reveals that the magnetic contribution dominates at low temperatures (T $<$ 15 K) and the magnetic heat capacity has similar values in bulk and 2D. Overall, due to the higher lattice heat capacity, 2D magnets have a slightly higher total heat capacity, which can reduce the enhancement of $\Delta T_{ad}$. To summarize our discussion so far, the 3D to 2D transition leads to (1) a lower Curie temperature and a reduced critical component, both of which contribute to an enhanced magnetocaloric entropy change $\Delta S_M$; (2) a lower required magnetic field to drive an MCE cycle; (3) a slightly increased lattice heat capacity that will reduce the enhancement in $\Delta T_{ad}$.


\subsection{\texorpdfstring{GdSi\textsubscript{2} -} A 2D Magnet with Exceptional MCE Performance}
\begin{figure}[!tb]
\includegraphics[width=\textwidth]{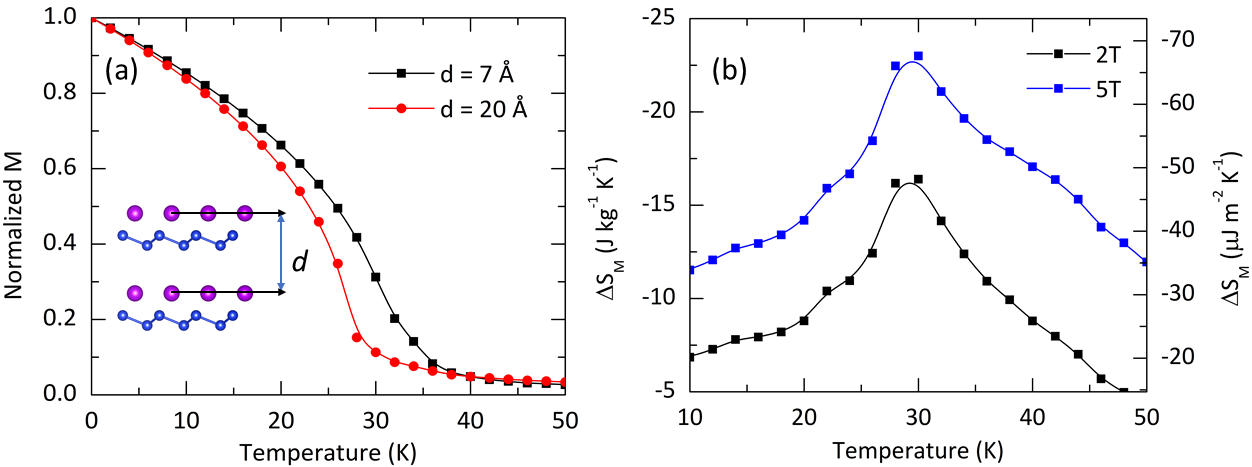}
\caption{\textbf{Magnetocaloric effect in 2D GdSi$_2$}: (a) Magnetization vs. temperature (M $\sim$ T) curves as a function of interlayer distance ($d$) for CrI$_3$. (b) The simulated magnetic entropy change Vs. temperature ($\Delta S_M \sim T$) curves for monolayer GdSi$_2$ under applied magnetic fields of 2T and 5T.}
\label{fig:fig4}
\end{figure}

Based on our understanding of the MCE effect in 2D magnets, we aim to discover new 2D magnets with promising MCE properties. In this section, we focus our attention on the MCE properties of a 2D rare-earth magnet GdSi$_2$, whose monolayer has been recently grown on a Si(111) surface.~\cite{tokmachev2018emerging} Each GdSi$_2$ monolayer can be conceptualized as a freestanding silicene monolayer forming a honeycomb lattice, with Gd atoms fully covering the central position of each hexagon of Si atoms. Notably, the GdSi$_2$ monolayer stabilizes as a ferromagnetic insulator despite the inherent antiferromagnetic ground state observed in the bulk GdSi$_2$ crystal.~\cite{tokmachev2018emerging, demirci2021magnetization} The inset of Fig.~\ref{fig:fig4}(a) provides an illustrative representation of the structural arrangement within the bilayer GdSi$_2$, comprising two buckled monolayers separated by an interlayer distance denoted as $d$.  In contrast to the monolayer cases of CrX$_3$, the MAE for the GdSi$_2$ monolayer, with an easy axis lying within the plane, is quantified at 0.175 meV. Due to the antiferromagnetism in bulk GdSi$_2$, we investigated the interlayer-distance-dependent magnetic properties by constructing a theoretical GdSi$_2$ bilayer. It is essential that the bilayer structure, optimized for the value of $d$, also maintains an antiferromagnetic (AFM) state and experiences a transition from AFM to ferromagnetic (FM) behavior when $d$  exceeds 6 \AA. Since ferromagnetic states usually have better MCE properties, we study the bilayers with the $d$ value larger than 7 \AA, at which point the ferromagnetic state achieves stability. 

By increasing $d$ from 7 \AA~ till the monolayer limit, we can then study the 3D to the 2D transition of the MCE properties in GdSi$_2$. Accordingly, we conduct simulations to determine the $T_C$ for the bilayer structure, comparing cases where $d$ equals 7 \AA~ and 20 \AA~[Fig.~\ref{fig:fig4}(a)]. From the simulations, we derive $T_C$ values of 32 K and 28 K for $d$ values of 7 \AA~ and 20 \AA~ in the GdSi$_2$ bilayer, respectively. We also simulate a free-standing monolayer of GdSi$_2$, which gives a $T_C$ of 25 K. These calculated $T_C$ values are slightly higher than those reported by Tokmachev et al.,~\cite{tokmachev2018emerging} potentially due to the growth of GdSi$_2$ layers on the Si(111) substrate, which can result in reduced magnetization at Gd sites. In addition, Demirci et al.~\cite{demirci2021magnetization} previously studied the free-standing GdSi$_2$ monolayer computationally, reporting a $T_C$ value of 122 K. This reported value markedly exceeds experimental measurements and could be influenced by the choice of Coulomb correlation parameters employed in the simulation. We note here that our simulations are conducted within the framework of PBE-DFT, as it is proven to be capable of predicting the physical properties of GdSi$_2$ and similar compounds~\cite{sanna2016rare, sanna2021spectroscopic, yang2016gadolinium, yun2006electronic} in various dimensionalities. Our ASD calculations further yield a $\Delta S_M$ value of $\sim$ -22.5 J kg$^{-1}$ K$^{-1}$ for the GdSi$_2$ monolayer under a magnetic field change of 5 T [Fig.~\ref{fig:fig4}(b)]. Remarkably, due to its high areal density, the $\Delta S_M$ value expressed in terms of the surface area is $\sim$ -65 $\mu$J m$^{-2}$ K$^{-2}$, marking it as the highest reported value for 2D magnets to date. 

\begin{figure}[!tb]
\includegraphics[width=\textwidth]{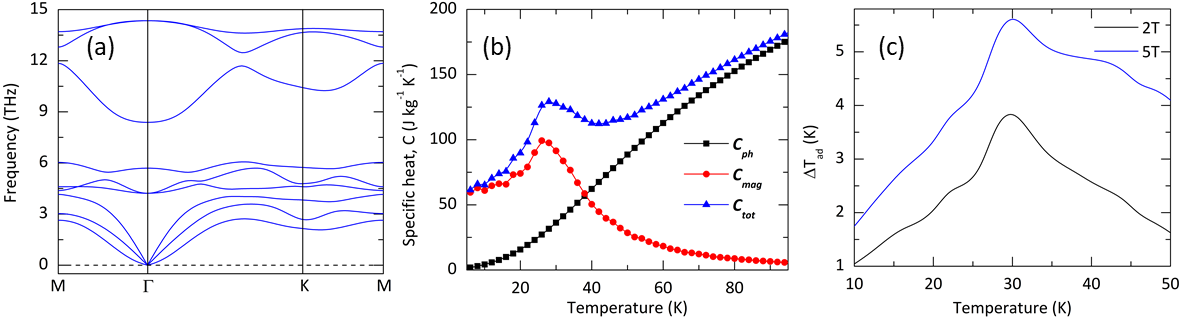}
\caption{\textbf{Heat capacity and adiabatic temperature change of monolayer GdSi$_2$:} (a) Phonon dispersion and (b) the total heat capacity ($C_{tot}$), comprising contributions from phonons ($C_{ph}$) and magnetism ($C_{mag}$), are presented for monolayer GdSi$_2$. (c) The calculated adiabatic temperature change ($\Delta T_{ad}$) of GdSi$_2$ for applied fields of 2T and 5T }
\label{fig:fig5}
\end{figure}

Figure~\ref{fig:fig5} shows the heat capacity and the adiabatic temperature change $\Delta T_{ad}$ of GdSi$_2$. The phonon dispersion of GdSi$_2$ is shown in Fig.~\ref{fig:fig5}(a), where the quadratic flexural phonon mode is present. Fig.~\ref{fig:fig5}(b) decomposes the total heat capacity to the magnetic and lattice contributions, indicating that the magnetic contribution dominates under 40 K. In addition, a notably high $\Delta T_{ad}$ value of $\sim$ 6.2 K is predicted for the monolayer of GdSi$_2$ under a magnetic field of 5 T [Fig.~\ref{fig:fig5}(c)]. The large $\Delta S_M$ and $\Delta T_{ad}$ of GdSi$_2$ in the 20 to 30 K range suggest its potential application in technologically important applications such as hydrogen liquefaction.~\cite{li2020understanding}

\section{Conclusion}
In summary, our investigation has revealed a stronger MCE observed in the vdW monolayers compared to their bulk counterparts, as analyzed through density functional theory and spin dynamics simulations. The lower critical exponent ($\beta$) value in the monolayer configuration leads to a sharper transition in the $\frac{dM}{dT}$ curve at $T_C$, resulting in a higher $\Delta S_M$. Our study highlights the significant role of out-of-plane exchange energy in reducing $T_C$ and enhancing MCE in the monolayers of the vdW layered magnets. This suggests that the vertical strain can effectively tune the MCE at desired temperatures. Additionally, the MCE in monolayers is more responsive to applied magnetic fields compared to 3D bulk materials. Due to the flexural phonon mode and generally lower phonon frequencies in 2D, however, the lattice contribution to the heat capacity in 2D magnets is higher, reducing the achievable adiabatic temperature change $\Delta T_{ad}$. Furthermore, our calculations predict giant MCE in recently synthesized GdSi$_2$ monolayers with a critical temperature (25 K) around hydrogen liquefaction temperature. Our study provides a systematic and general understanding of magnetocaloric properties in 2D ferromagnets and suggests that promising magnetocaloric materials for a wide range of application temperatures can be found in this group of novel magnetic materials.

\begin{acknowledgments}
We thank Dr. Amir Jahromi and Dr. Ali Kashani for their helpful discussions. This work is based on research supported by the National Aeronautics and Space Administration (NASA) under award number 80NSSC21K1812. Y.Q. also acknowledges support from the NSF Quantum Foundry via the Q-AMASE-i program under award number DMR-1906325 at the University of California, Santa Barbara (UCSB). This work used Stampede2 at Texas Advanced Computing Center (TACC) through allocation MAT200011 from the Advanced Cyberinfrastructure Coordination Ecosystem: Services \& Support (ACCESS) program, which is supported by National Science Foundation grants 2138259, 2138286, 2138307, 2137603, and 2138296. Use was also made of computational facilities purchased with funds from the National Science Foundation (award number CNS-1725797) and administered by the Center for Scientific Computing (CSC) at the University of California, Santa Barbara (UCSB). The CSC is supported by the California NanoSystems Institute and the Materials Research Science and Engineering Center (MRSEC; NSF DMR-2308708) at UCSB. 
\end{acknowledgments}

\bibliography{references.bib}

\end{document}